\documentclass{iopart}
\usepackage{graphicx}
\usepackage{amstext}

\eqnobysec

\begin{document}

\title{Quantum mass correction for the twisted kink}
\author{Michael Pawellek}
\address{Institut f\"ur Theoretische Physik III, Universit\"at Erlangen-N\"urnberg, Staudtstr.7, D-91058 Erlangen, Germany}
\ead{michi@theorie3.physik.uni-erlangen.de}
\begin{abstract}
We present an analytic result for the 1-loop quantum mass correction in semiclassical quantization for the twisted \(\phi^4\)-kink on \(S^1\) 
without explicit knowledge of the fluctuation spectrum. For this purpose we use the contour integral representation of the spectral zeta function. By
solving the Bethe ansatz equations for the \(n=2\) Lam\'e equation we obtain an analytic expression for the corresponding spectral discriminant. 
We discuss the renormalization issues of this model. An energetically preferred size for the compact space is finally obtained.
\end{abstract}
\section{Introduction}
It has been known for a long time, that on non-simply connected spaces there can exist besides the standard scalar fields
topological twisted fields \cite{Isha, Toms}. Recently there is renewed interest in phenomenological and theoretical aspects of this kind of field theory \cite{Jaff}, especially for the twisted version 
of the \(\phi^4\)-model with kink on \(S^1\) \cite{Muss,Saka}. This model possess some interesting features, e.g. there appears a critical radius \(R_0\) for the 
compactified dimension, so that for \(R<R_0\) a twisted kink solution does not exist. Even more, for \(R>R_0\) the twisted kink is energetically preferred 
compared to the constant field configuration. 

Compact spaces are also important in superstring theories, which are consistent only in ten space-time dimensions. If these theories describe the 
observed physical world, one has to explain why six space dimensions remain compactified and unobservable small. There are proposals that a Casimir energy 
with a nontrivial behaviour with respect to the size of the compact dimensions may play a significant role in their stabilization \cite{Gree}.

Assuming that in a (1+1)-dim. quantum field theory the radius \(R\) and a mass scale \(1/m\) are the only parameters with dimensions of length
(\(c=\hbar=1\)) then from dimensional considerations the ground state energy has the general property \(E(R)=f(r)/R\), where the scaling function \(f(r)\) 
only depends on the dimensionless parameter \(r=Rm\). These scaling functions contain information about 
the conformal field theory reached for \(r\to 0\), which is a UV-fixpoint \cite{Muss, Rava}.

To understand the quantum properties of the twisted \(\phi^4\)-model in the semiclassical regime, one has
to consider the 1-loop corrections to the ground state \cite{Dash}. For \(R<R_0\) the spectrum of the fluctuation equation can be 
found in \cite{Muss}. For \(R>R_0\) the ground state is the twisted kink and one has to quantize the small fluctuations in a spatial non-constant 
background, where the fluctuation equation is the \(n=2\) Lam\'e equation, which is a quasi-exactly solvable differential equation \cite{Turb, Arsc}.
This means that only a finite subset of the (anti-)periodic spectrum is exactly known. Therefore only approximate expressions for the mass of the kink for
\(R\sim R_0\) and \(R\to\infty\) were obtained so far \cite{Muss}.

In order to find the mass correction without explicit knowledge of the eigenvalues, we use the contour integral representation of spectral
zeta functions \cite{Bord, Kir1}, where 
only an implicit knowledge of the fluctuation spectrum, the spectral discriminant, is necessary to determine the 1-loop energy of a smooth 
background field configuration (for an early attempt in this direction see \cite{Bra1, Bra2}). This method was successfully applied to Casimir energy 
calculations (for a review see \cite{Kir1} or \cite{Bor2}) or to the evaluation of functional determinants \cite{Kir3, Kir4} .

We construct the spectral discriminant of the \(n=2\) Lam\'e equation in terms of Jacobi's elliptic functions \cite{Whit} (the  case \(n=1\) was 
solved in \cite{Bra1}, which is the fluctuation equation for the Sine-Gordon soliton on \(S^1\)) by solving a corresponding set of transcendental
equations. These equations has been known for a long time \cite{Whit} and are the Bethe ansatz equations for the \(n=2\) Lam\'e equation \cite{Take}, because the 
problem of solving a differential equation is shifted to the equivalent problem of solving certain transcendental equations. Although the case \(n=2\) 
was considered in \cite{Bra2} an explicit construction of the spectral discriminant appropriate for numerical evaluations was missing there.
Recently the \(n=2\) Lam\'e equation also appears in a model of Tachyon condensation in String theory \cite{Brit}

The renormalized expression for the 1-loop quantum mass of the twisted kink in the sector \(R>R_0\) obtained by this procedure interpolates continuously 
between the well known result for the ordinary kink of the \(\phi^4\)-model \cite{Bord,Raja} for \(R\to\infty\) and the ground state energy in the sector 
\(R<R_0\) \cite{Muss}. The physical energy which is the sum of the classical and renormalized 1-loop contributions develops a minimum as
function of \(R\), which indicates the existence of an energetically preferred radius \(R_{min}<R_0\).

\section{Twisted scalar field}
In this section we review the classical twisted kink solution in the \(\phi^4\)-theory \cite{Muss, Saka}. After that
we discuss the special properties of the fluctuation equation of the twisted kink, which is the \(n=2\) Lam\'e equation.

\subsection{Classical solutions}
We consider a self-interacting scalar field \(\phi(x)\) in two space-time dimensions with Lagrangian
\begin{equation}
 \mathcal{L}=\frac{1}{2}\partial_{\mu}\phi\partial^{\mu}\phi-V(\phi),\qquad V(\phi)=\frac{\lambda}{4}(\phi^2-\frac{m^2}{\lambda})^2
\end{equation}
with the spatial direction compactified with radius \(R\). By choosing antiperiodic conditions for the scalar field \(\phi(x+R)=-\phi(x)\).
the only allowed constant field configuration is \(\phi(x)=0\). 

In order to find \(x\)-dependent static solutions, one has to integrate the static field equation
\begin{equation}
 \frac{\mathrm{d}^2\phi}{\mathrm{d}x^2}-V'(\phi)=0
\end{equation}
twice. After solving an elliptic integral (for details see e.g. \cite{Muss,Saka}) one gets
\begin{equation}\label{eq:TwistKink}
 \phi_0(x)=\frac{m}{\sqrt{\lambda}}\sqrt{\frac{2k^2}{k^2+1}}\mathrm{sn}\left(\frac{mx}{\sqrt{k^2+1}},k\right),
\end{equation}
where \(\mathrm{sn}(x,k)\) is a Jacobi elliptic function. The elliptic modulus \(k\) depends on the Radius \(R\) of the compactified dimension:
\begin{equation}
 R=\frac{2}{m}\sqrt{1+k^2}\mathbf{K}(k),
\end{equation}
where \(\mathbf{K}(k)\) is the complete elliptic integral of the first kind.
The energy of this classical field configuration can be expressed in terms of complete elliptic integrals of first and second kind:
\begin{equation}\label{eq:TKEnergy}
   E(k)=\frac{m^3}{6\lambda}\frac{1}{(k^2+1)^{\frac{3}{2}}}\left[(k^2-1)(5+3k^2)\mathbf{K}(k)+8(k^2+1)\mathbf{E}(k)\right].
\end{equation}
Remembering the properties of \(\mathbf{K}(k)\), one sees that for \(k\to 1\) the spatial dimension becomes decompactified (\(R\to\infty\)) and the static solution (\ref{eq:TwistKink}) reduces to
the well known \(\phi^4\)-kink solution
\begin{equation}
 \phi_0(x)\stackrel{k\to 1}{\longrightarrow}\frac{m}{\sqrt{\lambda}}\tanh(\frac{m}{\sqrt{2}}x),
\end{equation}
while the energy (\ref{eq:TKEnergy}) becomes nothing else than the classical mass of the kink 
\begin{equation}
 E(k)\stackrel{k\to 1}{\to}\frac{2\sqrt{2}}{3}\frac{m^3}{\lambda}.
\end{equation}
For \(k\to 0\) the amplitude of the kink is forced to become zero. This happens at the critical radius 
\begin{equation}
 R_0=\frac{\pi}{m}.
\end{equation}
The value of the energy 
\begin{equation}
 E(k=0)=\frac{m^3}{4\lambda}\pi=\frac{m^4}{4\lambda}R_0
\end{equation}
matches at the point \(R=R_0\) with the energy of the constant field configuration \(\phi(x)=0\):
\begin{equation}
 E_{\phi=0}(R)=\frac{m^4}{4\lambda}R.
\end{equation}

\subsection{1-Loop fluctuations}
To investigate the stability of static solutions and for semiclassical quantization one has to expand the field in the Lagrangian into 
a static part and a fluctuating part \(\phi(x,t)=\phi_0(x)+e^{i\sqrt{\lambda}t}\chi(x)\) \cite{Raja}, where the small fluctuations have also to be antiperiodic
\begin{equation}
 \chi(x+R)=-\chi(x).
\end{equation}
For the determination of the fluctuation spectrum in the Minkowski vacuum without non-trivial boundary conditions, one has to expand about
\(\phi_0=\pm\frac{m}{\sqrt{\lambda}}\), which are the true vacuum states. The fluctuation equation is then given by
\begin{equation}\label{eq:KappaFluct}
 -\frac{\mathrm{d}^2}{\mathrm{d}x^2}\chi(x)=\kappa^2\chi(x)
\end{equation}
when introducing the momentum-like parameter \(\kappa^2=\lambda-2m^2\).
Thus the elementary excitations in this vacuum have a mass \(\sqrt{2}m\).

Every quantization about a non-trivial background \(\phi_0(x)\) will induce a potential \(U(x)\) in the fluctuation equation
compared to the Minkowski vacuum:
\begin{equation}\label{eq:MomentEigeneq}
 \left[-\frac{\mathrm{d}^2}{\mathrm{d}x^2}+U(x)\right]\chi(x)=\kappa^2\chi(x).
\end{equation}
Using \(\phi_0=0\) and \(\phi_0(x)\) given by (\ref{eq:TwistKink}) for \(R<R_0\) and \(R>R_0\) respectively, one can find the corresponding 
potentials as
\begin{eqnarray}
 U(x)=-3m^2,\;\;R<R_0,\nonumber\\
 U(x)=\frac{6m^2k^2}{1+k^2}\mathrm{sn}^2\left(\frac{mx}{\sqrt{1+k^2}},k\right)-3m^2,\;\;R>R_0.
\end{eqnarray}
For \(R<R_0\) the energy eigenvalues \(\lambda\) are trivially found as
\begin{equation}\label{eq:fluctspect}
 \lambda_n=\frac{(2n+1)^2}{R^2}\pi^2-m^2.
\end{equation}
In order to have a stable configuration, all eigenvalues \(\lambda_n\) have to be positive.
It is easily seen that this is the case only for \(R\leq\frac{\pi}{m}=R_0\).
It follows that for \(R>R_0\) the constant field configuration is unstable.

For \(R>R_0\) the fluctuation equation is the the \(n=2\) Lam\'e equation in Jacobian form
\begin{equation}\label{eq:LameFluct}
 \left[-\frac{\mathrm{d}^2}{\mathrm{d}\bar x^2}+n(n+1)k^2\mathrm{sn}^2(\bar x,k)\right]\chi(\bar x)=h\chi(\bar x),
\end{equation}
where 
\begin{equation}\label{eq:relation}
 h=\left(\frac{\kappa^2}{m^2}+3\right)(1+k^2)
\end{equation}
and \(\bar x=mx/\sqrt{1+k^2}\). The term \(n(n+1)k^2\mathrm{sn}^2(x)\) interpreted as a potential in a Schroedinger equation
is called finite-gap potential since the spectrum has \(n\) forbidden bands \cite{Novi}. The first five eigenvalues are known (with \(\bar\lambda=\lambda/m^2\)) \cite{Arsc}:
\begin{eqnarray}
 \fl \bar\lambda_{1,5}=1\pm2\frac{\sqrt{1-k^2(1-k^2)}}{k^2+1},\quad\bar\lambda_2=0,\quad\bar\lambda_3=\frac{3k^2}{1+k^2},\quad
 \bar\lambda_4=\frac{3}{1+k^2}.
\end{eqnarray}
Besides the lowest eigenvalue, these are the endpoints of the forbidden bands.
The corresponding eigenfunctions are Lam\'e polynomials (e.g. \(\chi_2(\bar x)=\mathrm{cn}(\bar x)\mathrm{dn}(\bar x)\)). 
Out of this five eigenfunctions only \(\chi_2\) and \(\chi_3\) have the required anti-periodicity. 
The eigenfunctions \(\chi_n(x)\) for \(n>5\) are called transcendental Lam\'e functions and can be written as infinite power series in Jacobi elliptic
functions. The corresponding eigenvalues \(\lambda_n\) as function of the elliptic modulus are not exactly known \cite{Erde,Arsc}.

\section{Spectral zeta functions}
In order to fix the notation we give in this section a short summary of zeta function regularization and the integral representation of spectral 
zeta functions \cite{Kir1, Kir3, Kir4}.

For the eigenvalue problem 
\begin{equation}\label{eq:DiffEq}
 D\phi(x,\lambda)=\lambda\phi(x,\lambda)
\end{equation}
with a second order differential operator \(D=-\partial_x^2+V(x)\) and properly chosen boundary conditions, the set of eigenvalues \(\{\lambda_i\}_{i\in\mathbf{N}}\)
is discrete and bounded from below. If (\ref{eq:DiffEq}) is a fluctuation equation obtained by a semiclassical expansion the 1-loop energy contribution 
to the classical solution is given by
\begin{equation}
 E_{1-loop}=\frac{1}{2}\sum_{n=0}^{\infty}\sqrt{\lambda_n}.
\end{equation}
In quantum field theories this expression is divergent and has to be regularized. In zeta function regularization one works with the spectral zeta 
function formally defined by
\begin{equation}\label{eq:Zetadef}
 \zeta_D(s)=\mu^{1+2s}\sum_{n=1}^{\infty}\lambda_n^{-s},
\end{equation}
with \(\mathrm{Re}(s)>s_0\), where \(s_0\) depends e.g. on the numbers of dimensions. The parameter \(\mu\) with dimension of mass is introduced in 
order that the energy has the correct dimension for all values of \(s\). One can show \cite{Seel, Eliz}, that \(\zeta_D(s)\) has a well-defined 
analytic continuation as meromorphic function over the whole complex plane \(s\in\mathrm{C}\). The 1-loop contribution to the energy of
a classical field configuration in zeta function regularization is then defined as the value of the analytic continuation of 
\(\zeta_D(s)\) at \(s=-\frac{1}{2}\):
\begin{equation}
 E_{1-loop}=\frac{1}{2}\zeta_D(-1/2).
\end{equation}
For renormalization we will apply the large mass subtraction scheme, which is widely used in Casimir energy calculations \cite{Bor2}. For a physical 
field with mass \(m\) one expects that all quantum fluctuations will be suppressed in the limit of large mass \(m\), because for a field with infinite 
mass the quantum fluctuations should vanish. So one expects that for \(m\to\infty\) there are no 1-loop corrections at all and a good 
renormalization condition is \cite{Bor4, Bor2}
\begin{equation}\label{eq:RenormCond}
 E_{ren}\to 0 ,\;\;\text{for}\;\;m\to\infty.
\end{equation}
With this prescription at hand one can identify and subtract the divergent (when \(s=-\frac{1}{2}\) is a pole of
\(\zeta_D(s)\)) contributions \(E_{div}(s)\) from \(E_{1-loop}(s)\) and the renormalized 
energy is then given by
\begin{equation}
 E_{ren}=\lim_{s\to-\frac{1}{2}}\left[E_{1-loop}(s)-E_{div}(s)\right].
\end{equation}
This can be achieved by introducing counterterms in the Lagrangian, which have to cancel the \(E_{div}\) parts \cite{Bor2}.

In principle this can be applied to our problem for \(R<R_0\) since we know the spectrum in this case (see (\ref{eq:fluctspect})). But 
later we will see that after the analytical continuation the \(m\to\infty\) limit is not directly accessible. 

As discussed in the last section the complete set of eigenvalues for the Lam\'e equation under (anti-)periodic boundary conditions is unknown, so 
representation (\ref{eq:Zetadef}) of the spectral zeta function is of no use for our problem in the case \(R>R_0\). We need a 
representation of the zeta function, where only an implicit knowledge of the eigenvalues is necessary (see \cite{Kir1, Bor2} or \cite{Bra1}). 
Assume we have a function \(\Delta(\lambda)\), whose zeros of n-th order are at the positions \(\lambda_i>0\) of the n-fold degenerate 
eigenvalues of the spectral problem under consideration:
\begin{equation}
 \Delta(\lambda)=0\;\; \Leftrightarrow\;\;\; \lambda\;\;\text{eigenvalue of }\;\; D.
\end{equation}
Such a function is called the spectral discriminant. 
Then one can write the spectral zeta function as a contour integral
\begin{equation}\label{eq:ZetaInt1}
 \zeta_D(s)=\frac{1}{2\pi i}\mu^{1+2s}\int_{\gamma}\mathrm{d}\lambda\lambda^{-s}R(\lambda),
\end{equation}
with resolvent \(R(\lambda)=\frac{\mathrm{d}}{\mathrm{d}\lambda}\ln\Delta(\lambda)\).
The integrand has a branch cut along the negative real axis and poles at the positions of the zeros of \(\Delta(\lambda)\).
The contour \(\gamma\) runs counterclockwise from \(+\infty+i\varepsilon\) to the smallest eigenvalue, crosses the real axis between zero and the
smallest eigenvalue and returns to \(+\infty-i\varepsilon\). Using the residue theorem, one obtains the original definition of the
zeta function (\ref{eq:Zetadef}).

A slight modification is needed, when the spectrum contains a zero mode \(\lambda_0=0\), which means that \(\Delta(\lambda)\) and \(R(\lambda)\)
have a zero and pole at \(\lambda=0\), respectively, the starting point of the branch cut. In this case one redefines 
\begin{equation}
 \Delta(\lambda)\to\frac{\Delta(\lambda)}{\lambda},\qquad R(\lambda)\to R(\lambda)-\frac{1}{\lambda}
\end{equation}
in previous equations. 

Depending on the behaviour of \(R(\lambda)\) at infinity, for suitable values of \(s\) the contour can now be deformed to lie just above and below 
the branch cut. One gets \cite{Bra1}
\begin{equation}\label{eq:ZetaIntLambda}
 \zeta_D(s)=-\frac{\sin(\pi s)}{\pi}\mu^{1+2s}\int_0^{\infty}\mathrm{d}\lambda\lambda^{-s}R(-\lambda).
\end{equation}
In terms of the momentum variable \(\kappa^2\) this expression is rewritten as
\begin{equation}\label{eq:ZetaIntKappa}
 \zeta_D(s)=-\frac{\sin(\pi s)}{\pi}\mu^{1+2s}\int_{\sqrt{2}m}^{\infty}\mathrm{d}\kappa(\kappa^2-2m^2)^{-s}\mathcal{R}(\kappa),
\end{equation}
with
\begin{equation}
 \mathcal{R}(\kappa)=\left.2\kappa R(-\lambda)\right|_{\lambda=\kappa^2-2m^2}.
\end{equation}
In deriving (\ref{eq:ZetaIntLambda}) we have changed \(\lambda\to -\lambda\), which corresponds to \(\kappa\to i\kappa\). So the correct substitution
of the integration variable in (\ref{eq:ZetaIntLambda}) is \(\kappa^2=\lambda+2m^2\) to get (\ref{eq:ZetaIntKappa}).

\section{Construction of the spectral discriminant}
In this section we construct the analytic expression for the spectral discriminant \(\Delta(h)\) for the standard \(n=2\) Jacobi form of the 
Lam\'e equation, which is given by
\begin{equation}\label{eq:JacobiLame}
 -\frac{\mathrm{d}^2f}{\mathrm{d}x^2}+6k^2\mathrm{sn}^2(x,k)f(x)=hf(x).
\end{equation} 
A check with (\ref{eq:MomentEigeneq}) and (\ref{eq:LameFluct}) shows, that \(h\) is related to \(\kappa^2\) by (see (\ref{eq:relation}))
\begin{equation}\label{eq:Shift2}
 h=\left(\frac{\kappa^2}{m^2}+3\right)(1+k^2).
\end{equation}
After we have found the discriminant for (\ref{eq:JacobiLame}) we only have to substitute (\ref{eq:Shift2}) into the found expression to get the 
discriminant we are physically interested in.

For second order differential operators \(-\mathrm{d}_x^2+V(x)\) with periodic potential \(V(x+R)=V(x)\) the discriminant \(\Delta(h)\)
is an entire function of \(h\) and has the general form \cite{Bra1, Bra2, Novi, Kohn, Smir}
\begin{equation}\label{eq:discriminant}
 \Delta(h)=2\cos(Rp(h))\pm 2,
\end{equation}
where the negative and positive signs correspond to periodic and antiperiodic solutions, respectively and 
\(p(h)\) is the quasi-momentum defined by
\begin{equation} 
 f_{h}(x+R)=e^{\pm ip(h)}f_{h}(x).
\end{equation}
The resolvent for the antiperiodic spectrum is then given by
\begin{equation}
 R(h)=-\tan\left(\frac{R}{2}p(h)\right)p'(h).
\end{equation}
The general solution for (\ref{eq:JacobiLame}) is given by \cite{Whit}
\begin{equation}\label{eq:GenSolution}
 f(x)=\frac{H(x+\alpha_1)H(x+\alpha_2)}{\Theta(x)^2}e^{-x(Z(\alpha_1)+Z(\alpha_2))},
\end{equation}
(\(H(x),\Theta(x)\) and \(Z(x)\) are the Jacobi eta, theta and zeta function respectively)
if the additional parameters \(\alpha_1,\alpha_2\) fulfil the following two transcendental equations:
\begin{eqnarray}\label{eq:TransEq}
 \mathrm{sn}(\alpha_1)\mathrm{cn}(\alpha_1)\mathrm{dn}(\alpha_1)+\mathrm{sn}(\alpha_2)\mathrm{cn}(\alpha_2)\mathrm{dn}(\alpha_2)=0,\nonumber\\
 (\mathrm{cn}(\alpha_1)\mathrm{ds}(\alpha_1)+\mathrm{cn}(\alpha_2)\mathrm{ds}(\alpha_2))^2-\mathrm{ns}^2(\alpha_1)-
 \mathrm{ns}^2(\alpha_2)=-h.
\end{eqnarray}
These equations are nowadays \cite{Take} called Bethe ansatz equations of the \(n=2\) Lam\'e potential.
The periodic properties of Jacobi's eta, theta and zeta functions \cite{Whit, Byrd} imply
\begin{equation}
 f(x+2\mathbf{K})=f(x)e^{2i\mathbf{K}p(\alpha_1,\alpha_2)}
\end{equation}
with the quasi-momentum 
\begin{equation}
 p(\alpha_1,\alpha_2)=iZ(\alpha_1)+iZ(\alpha_2).
\end{equation}
In order to find the dependence of the quasi momentum in terms of the eigenvalue parameter \(h\) we solve
the Bethe ansatz equations (\ref{eq:TransEq}). These equations can be written in terms of \(\mathrm{sn}\)-functions only:
\begin{eqnarray}
\fl 2k^2\mathrm{sn}^4\alpha_1-2(1+k^2)(\mathrm{sn}^2\alpha_1-\mathrm{sn}^2\alpha_2)-k^2\mathrm{sn}^2\alpha_1\mathrm{sn}^2\alpha_2-
 k^2\mathrm{sn}^4\alpha_2-h\>\mathrm{sn}^2\alpha_2+2=0,\nonumber\\
\fl 2k^2\mathrm{sn}^4\alpha_2+2(1+k^2)(\mathrm{sn}^2\alpha_1-\mathrm{sn}^2\alpha_2)-k^2\mathrm{sn}^2\alpha_1\mathrm{sn}^2\alpha_2-
 k^2\mathrm{sn}^4\alpha_1-h\>\mathrm{sn}^2\alpha_1+2=0.\nonumber
\end{eqnarray}
The solutions of this equations are found to be
\begin{eqnarray}\label{eq:Translos}
 \mathrm{sn}^2\alpha_1&=&\frac{4(1+k^2)-h}{6k^2}+\frac{1}{2k^2}\sqrt{g_2(k)-\frac{1}{3}(h-2(1+k^2))^2},\\
 \mathrm{sn}^2\alpha_2&=&\frac{4(1+k^2)-h}{6k^2}-\frac{1}{2k^2}\sqrt{g_2(k)-\frac{1}{3}(h-2(1+k^2))^2},
\end{eqnarray}
where
\begin{equation}
 g_2(k)=\frac{4}{3}(1-k^2(1-k^2)).
\end{equation}
Next we eliminate the dependence of the quasi-momentum on \(\alpha_1,\alpha_2\) in favour of \(h\)
\begin{eqnarray}\label{eq:Quasimoment}
\fl p(h)&=&iZ\left[\mathrm{sn}^{-1}\left(\sqrt{\frac{4(1+k^2)-h}{6k^2}+\frac{1}{2k^2}\sqrt{g_2(k)-
 \frac{1}{3}(h-2(1+k^2))^2}}\right)\right]+\nonumber\\
\fl & &+iZ\left[-\mathrm{sn}^{-1}\left(\sqrt{\frac{4(1+k^2)-h}{6k^2}-\frac{1}{2k^2}\sqrt{g_2(k)-\frac{1}{3}(h-2(1+k^2))^2}}\right)\right]
\end{eqnarray} 
and the spectral discriminant for the antiperiodic eigenfunctions is given by
\begin{equation}
 \Delta(h)=2\cos\left(2\mathbf{K}(k)p(h)\right)+2=4\cos^2\left(\mathbf{K}(k)p(h)\right).
\end{equation}
Because we are interested only in the zero points of this function, the prefactor of \(4\) is not essential in the following and can be omitted. 
The resolvent \(R(h)\) can then be written as
\begin{equation}\label{eq:Resolvent3}
 R(h)=-\mathbf{K}(k)\tan[\mathbf{K}(k)p(h)]p'(h),
\end{equation}
where the first derivative of the quasi-momentum is given by
\begin{equation}\label{eq:Quasimomentprime}
p'(h)=-\frac{i}{2}\frac{(h-\mu_1)(h-\mu_2)}{\sqrt{(h-h_1)(h-h_2)(h-h_3)(h-h_4)(h-h_5)}}
\end{equation}
with
\begin{eqnarray}
 \mu_{1,2}&=&\frac{3}{2}\frac{\mathbf{E}(k)}{\mathbf{K}(k)}+\frac{5}{2}k^2+1\pm\frac{3}{2}\sqrt{\frac{2}{3}g_2(k)+\left(\frac{\mathbf{E}(k)}{\mathbf{K}(k)}-
 \frac{(2-k^2)}{3}\right)^2},\\
 h_2&=&1+k^2,\;\;h_3=1+4k^2,\;\;h_4=4+k^2,\nonumber\\
 h_{5,1}&=&2(1+k^2)\pm2\sqrt{1-k^2(1-k^2)}
\end{eqnarray}
where \(h_i\) are the eigenvalues at the endpoints of the forbidden bands and \(\mu_i\) are the first two local extrema of \(\Delta(h)\),
which lie inside the two forbidden bands.

The quasi momentum \(p(h)\) and its derivative \(p'(h)\) are double-valued functions with branch points \(h_i, i=1,..,5\) and 
\(\infty\) and branch cuts along \((-\infty,h_1],[h_2,h_3], [h_4,h_5]\). For values of \(h\) on the
cuts one has \(p(h+i\varepsilon)=-p(h-i\varepsilon)\) for \(\varepsilon\to 0\) and therefore 
the functions \(\Delta(h)\) and \(R(h)\) are single valued and have no cuts in the complex plane. 

Although for the resolvent (\ref{eq:Resolvent3}) all five known eigenvalues are needed, it has only poles at points, which are the eigenvalues of 
the corresponding anti-periodic eigenfunction, whose sequence starts with \(h_2\) and \(h_3\).

\section{The 1-loop contributions}
In this section we derive the renormalized 1-loop contributions to the ground state of the twisted \(\phi^4\)-theory. First
we discuss the regularization of the energy in the sector \(R<R_0\) where only \(\phi=0\) is permitted and argue that the large mass renormalization 
condition (\ref{eq:RenormCond}) cannot be applied as usual. Then we consider the twisted kink sector \(R>R_0\) and find the renormalized 1-loop contribution
to its mass by using (\ref{eq:RenormCond}). Afterwards we go back to the sector \(R<R_0\) and use the condition that the renormalized energies in both 
sectors have to match for \(R=R_0\).

\subsection{Regularization in the sector \(R<R_0\)}
We will find two equivalent expressions for the regularized ground state energy in this sector. The first one is obtained by analytical continuation by a 
binomial expansion of the original expression for the zeta function \cite{Muss,Kir2}. The second one is the renormalized integral representation of 
the 1-loop energy for \(R<R_0\) and is a new result of this work. In \cite{Muss} the case \(R\to 0\) was discussed.

We start with the fluctuation spectrum (\ref{eq:fluctspect}) for \(R<R_0\) which is given by (\ref{eq:fluctspect}). The corresponding spectral zeta function 
\begin{equation}\label{eq:Regzeta}
 \fl \zeta_D(s)=\mu^{1+2s}\sum_{n=-\infty}^{\infty}\lambda_n^{-s}=\mu^{1+2s}\sum_{n=-\infty}^{\infty}\left[\left(\frac{(2n+1)\pi}{R}\right)^2-
 m^2\right]^{-s}.
\end{equation}
converges for \(\mathrm{Re}(s)>\frac{1}{2}\). As discussed in section 3, we need for the 1-loop energy the value
of the the zeta function at \(s=-\frac{1}{2}\), which lies outside the convergence region. The analytical continuation by a
Mellin transformation is not possible because of the negative sign in front of \(m^2\). What we can do is a continuation by a
binomial expansion. Following the steps given in Appendix A one finds (see also \cite{Kir2} for a discussion of the series with \(\lambda_n=
(n+c)^2+m^2)\)
\begin{equation}
 \fl \zeta_D(s)=2\mu\left(\frac{R\mu}{2\pi}\right)^{2s}\sum_{k=0}^{\infty}\frac{\Gamma(s+k)}{k!\Gamma(s)}\left(\frac{mR}{2\pi}\right)^{2k}
 (2^{2s+2k}-1)\zeta(2s+2k),
\end{equation}
where \(\zeta(x)\) is the Riemann zeta function.
This is the analytical continuation of \(\zeta_D(s)\) to the region \(\mathrm{Re}(s)<\frac{1}{2}\). One sees immediately from
the pole of the Riemann zeta function at \(x=1\), that \(\zeta_D(s)\) has poles at \(s=\frac{1}{2}-k\), \(k\in\mathbf{N}_0\). We are 
particularly interested in the singularity at 
\(s=-\frac{1}{2}\). With \(s=-\frac{1}{2}+\varepsilon\), we can make a Laurent expansion around \(\varepsilon=0\) for \(\zeta_D(s)\) to get the
divergence explicit. The result reads:
\begin{eqnarray}\label{eq:BinomialExpand}
 \fl \zeta_D(-\frac{1}{2})=-\frac{2\pi}{R}\zeta(-1)+\frac{m^2R}{2\pi}\left[\left.-\frac{1}{2s+1}\right|_{s=-\frac{1}{2}}+1-\gamma-\ln\left(
 \frac{2R\mu}{\pi}\right)\right]+\nonumber\\
 +\frac{4\pi}{R}\sum_{n=2}^{\infty}\frac{\Gamma(n-\frac{1}{2})}{n!\Gamma(-\frac{1}{2})}\left(\frac{mR}{2\pi}\right)^{2n}(2^{2n-1}-1)\zeta(2n-1),
\end{eqnarray}
where \(\gamma\) is the Euler constant.

At this point one wants to use the renormalization prescription (\ref{eq:RenormCond}). If we naively do this, we have to discard the term with squared
brackets. But this is not correct. We cannot apply the renormalization prescription here, because this expression is only valid for
\(mR<\pi\). After renormalization in the sector \(R>R_0\) we will revisit expression (\ref{eq:BinomialExpand}). 

We turn to the integral representation of \(\zeta_D(s)\), described in section 3.
For \(R<R_0\) the complete set of eigenfunctions \(\lambda_n\) is known and thus the corresponding spectral discriminant \(\Delta(\lambda)\): 
\begin{equation}\label{eq:eigenval}
 \lambda_n=\left(\frac{(2n+1)\pi}{R}\right)^2-m^2\;\;\;\;\Leftrightarrow\;\;\;\Delta(\lambda)=\cos^2\left(\frac{R}{2}\sqrt{\lambda+m^2}\right).
\end{equation}
The integral representation is given by (\ref{eq:ZetaIntLambda}) with
\begin{equation} 
 R(-\lambda)=-\frac{R}{2}\frac{\tanh\left(\frac{R}{2}\sqrt{\lambda-m^2}\right)}{\sqrt{\lambda-m^2}}.
\end{equation}
In this expression we have already deformed the integration contour from the poles on the positive real axis to the
branch cut along the negative real axis. This is valid for \(\frac{1}{2}<\mathrm{Re}(s)<1\) and \(mR<\pi\). The restriction \(mR<\pi\) is necessary
since for fixed radius \(R\) the first eigenvalues (\ref{eq:eigenval}) become negative when \(mR\) becomes larger than \(\pi\) and
the corresponding poles of \(R(\lambda)\) move into the branch cut which makes the integral representation invalid.

For better comparison with other results in the literature we finally switch to the momentum integration variable \(\kappa\):
\begin{equation}\label{eq:intrepzeta}
 \zeta_D(s)=-\mu^{1+2s}\frac{\sin(\pi s)}{\pi}\int_{\sqrt{2}m}^{\infty}\mathrm{d}\kappa(\kappa^2-2m^2)^{-s}\mathcal{R}(\kappa),
\end{equation} 
with 
\begin{equation}
 \mathcal{R}(\kappa)=-\frac{R\kappa\tanh\left(\frac{R}{2}\sqrt{\kappa^2-3m^2}\right)}{\sqrt{\kappa^2-3m^2}}.
\end{equation}
The limitation \(\frac{1}{2}<\mathrm{Re}(s)\) follows from the divergent behaviour of the integral for \(\kappa\to\infty\).
The asymptotic expansion of \(\mathcal{R}(\kappa)\) for \(\kappa\to\infty\) is found to be given by
\begin{equation}
 \mathcal{R}(\kappa)\to -R-\frac{3m^2R}{2\kappa^2}+\mathcal{O}(\kappa^{-4}).
\end{equation}
Inserting these first two terms of the asymptotic expansion separately into (\ref{eq:intrepzeta}) one finds two terms where simple poles are hidden for \(s=-\frac{1}{2}\) as can 
be seen after a Laurent expansion for \(s=-\frac{1}{2}+\varepsilon\) around \(\varepsilon=0\):
\begin{eqnarray}\label{eq:DivTerms}
 \fl \lim_{s\to-\frac{1}{2}}E_{div}^{(1)}(s)=\left.R\frac{\sin(\pi s)}{2\pi}\mu^{1+2s}\int_{\sqrt{2}m}^{\infty}\mathrm{d}\kappa(\kappa^2-2m^2)^{-s}
 \right|_{s\to-\frac{1}{2}}=\nonumber\\ {} \nonumber\\
 =\frac{m^2R}{4\pi}\left[\left.\frac{2}{2s+1}\right|_{s\to-\frac{1}{2}}-1+2\ln\left(\frac{\sqrt{2}\mu}{m}\right)\right],\nonumber\\
 \fl \lim_{s\to-\frac{1}{2}}E_{div}^{(2)}(s)=\left.\frac{3m^2R}{2}\frac{\sin(\pi s)}{2\pi}\mu^{1+2s}\int_{\sqrt{2}m}^{\infty}\mathrm{d}\kappa
 \frac{(\kappa^2-2m^2)^{-s}}{\kappa^2}\right|_{s\to-\frac{1}{2}}=\nonumber\\ {} \nonumber\\
  =\frac{3m^2R}{4\pi}\left[\left.-\frac{1}{2s+1}\right|_{s\to-\frac{1}{2}}+1-\ln\left(\frac{\sqrt{2}\mu}{m}\right)\right].
\end{eqnarray}
Again, we have made the divergences explicit, but cannot apply immediately the renormalization condition (\ref{eq:RenormCond}), since these
results were derived from an expression valid for \(mR<\pi\). We will revisit (\ref{eq:intrepzeta}) after renormalization in the sector \(R>R_0\).

\subsection{Regularization and renormalization in the sector \(R>R_0\)}

Now we come to the interesting case \(R>R_0\). In this sector we have to use the integral representation of the spectral
zeta function, because only five eigenvalues of a discrete infinity set are exactly known.
The relation between the physical eigenvalues \(\kappa_i^2\) of section 2 and the mathematical eigenvalues \(h_i\) of section 4 is
\begin{equation}
 \kappa^2=m^2\left(-3+\frac{h}{1+k^2}\right).
\end{equation}
With the results of our work in section 4 (see (\ref{eq:Quasimoment}) and (\ref{eq:Quasimomentprime})) we can immediately write down 
the integral representation for \(\frac{1}{2}<\mathrm{Re}(s)<1\) of our spectral zeta function as
\begin{equation}\label{eq:intrepzeta2}
 \fl \zeta_D(s)=-\mu^{1+2s}\frac{\sin(\pi s)}{\pi}\int_{\sqrt{2}m}^{\infty}\mathrm{d}\kappa(\kappa^2-2m^2)^{-s}\left(\mathcal{R}(\kappa)+
 \frac{2\kappa}{\kappa^2-2m^2}\right)
\end{equation} 
with
\begin{equation}\label{eq:res}
 \fl \mathcal{R}(\kappa)=-R\kappa\tanh\left(\frac{R}{2}\tilde p(\kappa)\right)\frac{(\kappa^2+\mu_1)(\kappa^2+\mu_2)}
 {\sqrt{(\kappa^2+\kappa_1^2)(\kappa^2+\kappa_2^2)(\kappa^2+\kappa_3^2)(\kappa^2+\kappa_4^2)(\kappa^2+\kappa_5^2)}}.
\end{equation}
The quasi-momentum \(\tilde p(\kappa)\) is given by (we have set \(p(\kappa)=i\tilde p(\kappa)\)) 
\footnotesize\begin{eqnarray}
 \fl \tilde p(\kappa)=\frac{m}{\sqrt{1+k^2}}\left\{Z\left[\mathrm{sn}^{-1}\left[\sqrt{\frac{(1+k^2)(1+\frac{\kappa^2}{m^2})}{6k^2}+\frac{1}{2k^2}\sqrt{g_2(k)-
 \frac{1}{3}(1-\frac{\kappa^2}{m^2})^2(1+k^2)^2}}\right],k\right]\right.+\nonumber\\
 \fl \left.+Z\left[-\mathrm{sn}^{-1}\left[\sqrt{\frac{(1+k^2)(1+\frac{\kappa^2}{m^2})}{6k^2}-\frac{1}{2k^2}\sqrt{g_2(k)-
 \frac{1}{3}(1-\frac{\kappa^2}{m^2})^2(1+k^2)^2}}\right],k\right]\right\},
\end{eqnarray}\normalsize
as is obtained from (\ref{eq:Quasimoment}) by the shift (remembering the change \(\kappa\to i\kappa\) after moving the contour)
\begin{equation}
 \lambda=\left(-\frac{\kappa^2}{m^2}+3\right)(1+k^2).
\end{equation}
The five known eigenvalues are given by
\begin{eqnarray}\label{eq:eigenvalue}
 \kappa_1^2=-2m^2,\;\;\kappa_2^2=\frac{k^2-2}{1+k^2}m^2,\;\;\kappa_3^2=\frac{1-2k^2}{1+k^2}m^2,\nonumber\\
 \kappa_{4,5}^2=\left(-1\pm\frac{2}{1+k^2}\sqrt{1-k^2(1-k^2)}\right)m^2.
\end{eqnarray}
The additional term in the integrand of (\ref{eq:intrepzeta2}) is necessary in order to eliminate the pole at \(\lambda=0\) before deforming the contour 
in the defining expression (\ref{eq:ZetaInt1}).
The first two local extremal points of the spectral discriminant are
\begin{equation}\label{eq:localex}
 \fl \mu_{1,2}=\frac{m^2}{2(1+k^2)}\left(3\frac{\mathbf{E}(k)}{\mathbf{K}(k)}-(4+k^2)\pm 3\sqrt{\frac{2}{3}g_2(k)+
 \left(\frac{\mathbf{E}(k)}{\mathbf{K}(k)}-\frac{2-k^2}{3}\right)^2}\right).
\end{equation}
The coefficients in the asymptotic expansion of \(\mathcal{R}(\kappa)\) for \(\kappa\to\infty\) can be written in terms of polynomials 
in \(\kappa_i^2\) and \(\mu_j\):
\begin{equation}
 \mathcal{R}(\kappa)\to r_0+\frac{r_1}{\kappa^2}+\mathcal{O}(\kappa^{-4}),\;\kappa\to\infty
\end{equation}
with
\begin{eqnarray}\label{eq:asymcoeff}
 r_0=-R=-\frac{2}{m}\sqrt{1+k^2}\mathbf{K}(k),\nonumber\\
 \fl r_1=-R\left[\mu_1+\mu_2-\frac{1}{2}\sum_{n=1}^5\kappa_n^2\right]=-\frac{3m}{\sqrt{k^2+1}}\left[(k^2-1)\mathbf{K}(k)+2\mathbf{E}(k)\right],
\end{eqnarray}
where we have used (\ref{eq:eigenvalue}) and (\ref{eq:localex}). Inserting this asymptotic form back into (\ref{eq:intrepzeta2}) and
making a Laurent expansion for \(s=-\frac{1}{2}+\varepsilon\) around \(\epsilon=0\) we find
\begin{eqnarray}\label{eq:divterms}
 \lim_{s\to-\frac{1}{2}}E_{div}^{(1)}(s)=-\frac{r_0m^2}{4\pi}\left[\left.\frac{2}{2s+1}\right|_{s\to-\frac{1}{2}}-1+
2\ln\left(\frac{\sqrt{2}\mu}{m}\right)\right],\nonumber\\
 \lim_{s\to-\frac{1}{2}}E_{div}^{(2)}(s)=-\frac{r_1}{2\pi}\left[-\left.\frac{1}{2s+1}\right|_{s\to-\frac{1}{2}}+1-
 \ln\left(\frac{\sqrt{2}\mu}{m}\right)\right].
\end{eqnarray}
Applying the large mass subtraction condition (\ref{eq:RenormCond}), we have to discard these terms completely:
\begin{equation}
 E_{ren}=\lim_{s\to-\frac{1}{2}}\left[E_{1-loop}(s)-E_{div}^{(1)}(s)-E_{div}^{(2)}(s)\right].
\end{equation}
The zero mode cancelling term present in (\ref{eq:intrepzeta2}) becomes zero in the used regularization for \(s\to-\frac{1}{2}\) and needs no
further subtraction.
We get as final result for the 1-loop contribution to the energy of the twisted kink in the sector \(R>R_0\) 
\begin{equation}\label{eq:FinalForm}
 E_{ren}(k)=\frac{1}{2\pi}\int_{\sqrt{2}m}^{\infty}\mathrm{d}\kappa\sqrt{\kappa^2-2m^2}\left[\mathcal{R}(\kappa)-r_0-\frac{r_1}{\kappa^2}\right],
\end{equation}
where \(\mathcal{R}(\kappa)\) and \(r_0\), \(r_1\) are given by (\ref{eq:res}) and (\ref{eq:asymcoeff}).
Equation (\ref{eq:FinalForm}) is the main result of this work. It gives the renormalized 1-loop energy of the twisted kink depending implicit on 
the Radius \(R\) by the elliptic modulus \(k\). Before we discuss the numerical evaluation of the remaining integral we will carry out the
renormalization in the sector \(R<R_0\).

\subsection{Renormalization in the sector \(R<R_0\)}
We have seen in section 5.1 that the large mass renormalization condition (\ref{eq:RenormCond}) cannot applied to (\ref{eq:intrepzeta}) for \(R<R_0\), 
but now we have a renormalized result for the energy for \(R>R_0\) and a natural renormalization condition for \(R<R_0\) is that the renormalized 
energy for \(R<R_0\) has to match at \(R=R_0\) the renormalized energy for \(R>R_0\):
\begin{equation}
 E_{ren,R<R_0}(R)\to E_{ren,R>R_0}(R_0) \text{ for } R\to R_0.
\end{equation}
With the results of Appendix B the quasi-momentum for \(R>R_0\) reduces to 
\begin{equation}
 \tilde p(\kappa)\to -\sqrt{\kappa^2-3m^2}
\end{equation}
for \(k\to 0\) which means \(R\to R_0\).
The eigenvalues \(\kappa_i^2\) and the extremal points \(\mu_i\) in this limit are
\begin{equation}
 \kappa_1^2,\kappa_2^2,\mu_2\to -2m^2,\;\;\;\kappa_3^2,\kappa_4^2,\mu_1\to m^2,\;\;\;\kappa_5^2\to -3m^2,\;\;\;k\to 0.
\end{equation}
With these results we find for (\ref{eq:res}) in this limit  
\begin{equation} 
 \mathcal{R}(\kappa)\to -\frac{R_0\kappa\tanh\left(\frac{R_0}{2}\sqrt{\kappa^2-3m^2}\right)}{\sqrt{\kappa^2-3m^2}},\;\;k\to 0.
\end{equation}
By noting that in the limit \(k\to 0\) the asymptotic coefficients (\ref{eq:asymcoeff}) become 
\begin{equation} 
 r_0\to -\frac{\pi}{m}=-R_0,\;\;\; r_1\to-\frac{3m\pi}{2}=-\frac{3m^2R_0}{2},
\end{equation}
it is seen that the subtraction terms (\ref{eq:divterms}) coincide with the divergent terms (\ref{eq:DivTerms}) at \(R=R_0\). Using now
the improved renormalization condition for \(R<R_0\) we see that we have to subtract (\ref{eq:DivTerms}) completely from (\ref{eq:intrepzeta})
and get
\begin{equation}\label{eq:RenEnergyInt}
 \fl E_{ren}(R)=-\frac{R}{2\pi}\int_{\sqrt{2}m}^{\infty}\mathrm{d}\kappa\sqrt{\kappa^2-2m^2}\left[\frac{\kappa\tanh\left(\frac{R}{2}
 \sqrt{\kappa^2-3m^2}\right)}{\sqrt{\kappa^2-3m^2}}-1-\frac{3m^2}{2\kappa^2}\right],
\end{equation}
which matches exactly the expression (\ref{eq:FinalForm}) when setting \(k=0\) and \(R=R_0\).

Since the subtraction terms for \(R<R_0\) are now identified as (see (\ref{eq:DivTerms}))
\begin{equation}
 E_{div}^{(1)}+E_{div}^{(2)}=\frac{m^2R}{4\pi}\left[\left.-\frac{1}{2s+1}\right|_{s\to-\frac{1}{2}}+2-\ln\left(\frac{\sqrt{2}\mu}{m}\right)\right].
\end{equation}
they have to be subtracted also from the expression of the analytically continued zeta function (\ref{eq:BinomialExpand}) obtained by binomial expansion. 
We get
\begin{eqnarray}\label{eq:RenEnergyBin}
 E_{ren}(R)=-\frac{\pi}{R}\zeta(-1)-\frac{m^2R}{4\pi}\left[1+\gamma+\ln\left(\frac{\sqrt{2}mR}{\pi}\right)\right]+\nonumber\\
 +\frac{2\pi}{R}\sum_{n=2}^{\infty}\frac{\Gamma(n-\frac{1}{2})}{n!\Gamma(-\frac{1}{2})}\left(\frac{mR}{2\pi}\right)^{2n}(2^{2n-1}-1)\zeta(2n-1).
\end{eqnarray}
Equation (\ref{eq:RenEnergyInt}) and (\ref{eq:RenEnergyBin}) are representations of the same function \(E_{ren}(R)\) valid for \(mR<\pi\).  
As a byproduct we have therefore obtained the following interesting identity
\begin{eqnarray}
 \fl \int_{\sqrt{2}}^{\infty}\mathrm{d}x\sqrt{x^2-2}\left[\frac{x\tanh\left(\frac{r}{2}\sqrt{x^2-3}\right)}{\sqrt{x^2-3}}-1-
 \frac{3}{2x^2}\right]=\frac{2\pi^2}{r^2}\zeta(-1)+\nonumber\\
 \fl +\frac{1}{2}\left[1+\gamma+\ln\left(\frac{\sqrt{2}r}{\pi}\right)\right]+
 \frac{2\sqrt{\pi}}{r^2}\sum_{n=2}^{\infty}\frac{\Gamma(n-\frac{1}{2})}{n!}\left(\frac{r}{2\pi}\right)^{2n}(2^{2n-1}-1)\zeta(2n-1),
\end{eqnarray}
where we have set \(\kappa=mx\) and \(r=Rm\).
 
\subsection{The limit \(R\to\infty\)}
For \(k\to 1\) the twisted kink reduces to the standard \(\phi^4\)-kink and (\ref{eq:FinalForm}) has to reproduce the standard mass correction 
formulas of the literature. In this limit the five eigenvalues \(\kappa_i\) and extremal points \(\mu_i\) of the spectral discriminant are
\begin{equation}
 \kappa_1^2,\kappa_5^2,\mu_2\to -2m^2,\;\;\;\kappa_2^2,\kappa_3^2,\mu_1\to -\frac{m^2}{2},\;\;\;\kappa_4^2\to 0,\;\;\;k\to 1.
\end{equation}
For \(R\to\infty\) our result (\ref{eq:FinalForm}) has to match with the kink mass calculated via the phase shift in \cite{Bord}:
\begin{equation}\label{eq:KinkEnergy}
 E_{ren}=\sum_{n=1}^2\frac{1}{2\pi}\int_{\sqrt{2}m}^{\infty}\mathrm{d}\kappa\sqrt{\kappa^2-2m^2}\frac{\mathrm{d}}{\mathrm{d}\kappa}
 \left[\delta(\kappa)-2\frac{\tilde\kappa_n}{\kappa}\right]
\end{equation}
with the derivative of the phase shift given by
\begin{equation}
 \frac{\mathrm{d}}{\mathrm{d}\kappa}\delta(\kappa)=\frac{\mathrm{d}}{\mathrm{d}\kappa}\sum_{n=1}^2\ln\frac{\kappa+\tilde\kappa_n}{\kappa-\tilde\kappa_n}=
 -2\frac{(\tilde\kappa_1+\tilde\kappa_2)(\kappa^2-\tilde\kappa_1\tilde\kappa_2)}{(\kappa^2-\tilde\kappa_1^2)(\kappa^2-\tilde\kappa_2^2)}
\end{equation}
with
\begin{equation}
 \tilde\kappa_1=\frac{m}{\sqrt{2}},\;\;\tilde\kappa_2=\sqrt{2}m.
\end{equation}
By comparison of (\ref{eq:FinalForm}) in the limit \(k\to 1\) with (\ref{eq:KinkEnergy}) we find the identity
\begin{equation}
 \lim_{k\to 1}\left[\mathcal{R}(\kappa)+\frac{2}{m}\sqrt{1+k^2}\mathbf{K}(k)\right]=-\frac{3\sqrt{2}m(\kappa^2-m^2)}{(\kappa^2-\frac{m^2}{2})
 (\kappa^2-2m^2)},
\end{equation}
which we have confirmed numerically.

\subsection{Comment on Renormalization schemes}
The physical results have to be independent of the renormalization scheme. 
In the perturbative approach to renormalizable quantum field theories the renormalization of the n-point functions is obtained by
counter terms (see \cite{Grah} for a discussion of this point in context of the Casimir effect). In \(\phi^4\) theory in (1+1) 
dimensions only the 2-point function requires infinite renormalization at one loop by the condition of vanishing tadpole Feynman 
graphs \cite{Raja}:
\begin{equation}
 \delta m^2=\frac{3\lambda}{4\pi}\int_{-\infty}^{\infty}\frac{\mathrm{d}k}{\sqrt{k^2+2m^2}}.
\end{equation}
In zeta function regularization this becomes
\begin{equation}
 \delta m^2(s)=\mu^{-1+2s}\frac{3\lambda}{4\pi}\int_{-\infty}^{\infty}\mathrm{d}k(k^2+2m^2)^{-s},
\end{equation}
and after analytic continuation on gets for \(s=\frac{1}{2}+\varepsilon\) and \(\varepsilon\to 0\):
\begin{equation}
 \delta m^2=\frac{3\lambda}{2\pi}\left[\frac{1}{2\varepsilon}+\ln\left(\frac{\sqrt{2}\mu}{m}\right)\right].
\end{equation}
The renormalization of the kink mass in the case of the infinite line consists of \cite{Raja}
\begin{equation}
 E=E_{cl}+E_{1-loop}-E_{vac}-E_{c.t},
\end{equation}
where \(E_{vac}\) is the divergent vacuum energy and \(E_{c.t}\) is given by (after regularization)
\begin{equation}
 E_{c.t.}=-\frac{\sqrt{2}m}{\lambda}\delta m^2=-\frac{3m}{\sqrt{2}\pi}\left[\frac{1}{2\varepsilon}+
 \ln\left(\frac{\sqrt{2}\mu}{m}\right)\right].
\end{equation}
In the previous section we used the large mass renormalization which results in
\begin{equation}
 E=E_{cl}+E_{1-loop}-E_{div}^{(1)}-E_{div}^{(2)}
\end{equation}
By (5.19) in the limit \(k\to 1\) one finds  
\begin{equation}
 E_{c.t}=E_{div}^{(2)}-\frac{3m}{\sqrt{2}\pi},\qquad E_{v}=E_{div}^{(1)}+\frac{3m}{\sqrt{2}\pi}
\end{equation}
With this identification the large mass renormalization scheme used in this work is consistent with the standard n-point function renormalization
prescription (see also \cite{Bor5}).

\section{Discussion of numerical evaluations and physical implications}

\begin{figure}
 \includegraphics[scale=0.7]{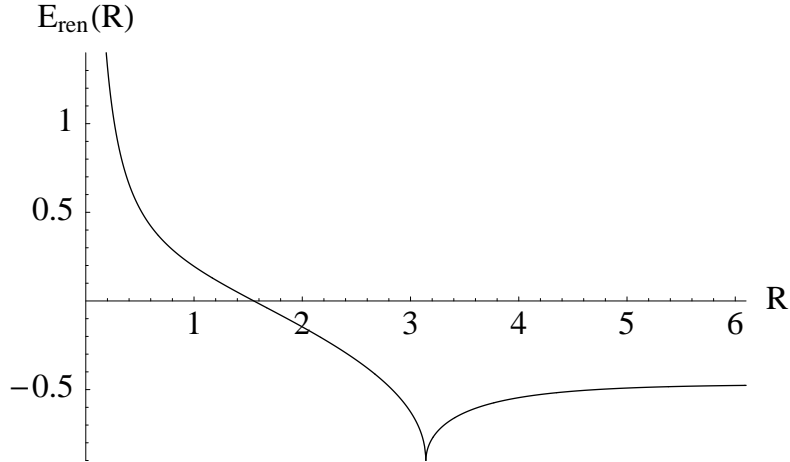}
 \caption{The renormalized 1-loop energy \(E_{ren}(R)\) with \(m=1,\lambda=0.1\)}
\end{figure}
\begin{figure}
 \includegraphics[scale=0.7]{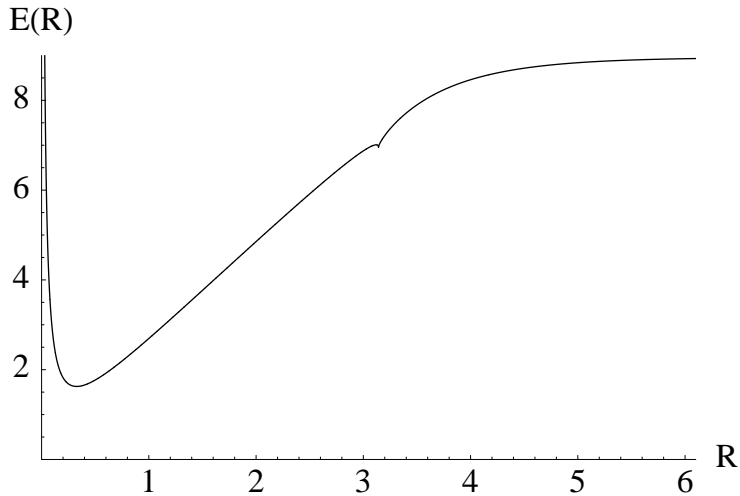}
 \caption{The physical energy \(E(R)\) with \(m=1,\lambda=0.1\)}
\end{figure}

In the following we present some numerical evaluations of our formulas derived in the last section. First we have to state the region 
of validity of our semiclassical quantization. In units where \(\hbar=1\), the dimensionless
expansion parameter is \(\lambda/m^2\) \cite{Raja}. Our results are valid as long as \(\lambda/m^2\ll 1\), we choose therefore 
\(\lambda=0.1\) and plotted the energy in units of \(m\). Although only 
\begin{equation}\label{eq:PhysicalEn}
 E(R)=E_{cl}(R)+E_{ren}(R)
\end{equation}
is the physical measurable quantity we have plotted \(E_{ren}(R)\) for its own. Figure 1 shows the 1-loop contribution \(E_{ren}(R)\) given by 
(\ref{eq:RenEnergyInt}) for \(R<R_0\) and (\ref{eq:FinalForm}) for \(R>R_0\). For small \(R\) this contribution behaves like the Casimir energy of a free 
massless scalar field with anti-periodic boundary conditions as can also 
be seen from (\ref{eq:RenEnergyBin}): 
\begin{equation}\label{eq:approx}
 E_{ren}(R)\to -\frac{\pi}{R}\zeta(-1),\;\;\;R\to 0.
\end{equation}

Before \(R\) approaches the critical radius \(R_0\) the function develops a turning point 
and goes to negative energy values. At the critical radius \(R_0\) a cusp appears. 

For \(R\to\infty\) \(E_{ren}(R)\) approaches the value of the 1-loop energy of the standard kink \cite{Bord, Raja}:
\begin{equation}
 E_{ren}\to \left(\frac{1}{2\sqrt{6}}-\frac{3}{\sqrt{2}\pi}\right)m=-0.4711m, k\to 1.
\end{equation}

Figure 2 shows the physical energy (\ref{eq:PhysicalEn}) of the \(\phi=0\) configuration in the sector \(R<R_0\) and the twisted kink for \(R>R_0\). 
Since \(R_0\approx 1/m\) one expects additional higher loop fluctuations
which travel around the compact dimension and may significantly contribute and smooth out the cusp in an all loop result.
Further we see that there exist an energetically preferred radius \(R_{min}<R_0\). 
Using the approximation (\ref{eq:approx}) the stabilization radius is given by
\begin{equation}\label{eq:minradius}
 R_{min}\approx\frac{1}{m}\sqrt{\frac{\lambda}{m^2}}\; (=0.31 \text{ for } m=1, \lambda=0.1).
\end{equation}
The stabilization radius exist since when we are going to smaller radius \(R\) the classical part of (\ref{eq:PhysicalEn}) becomes 
less and less important compared to the growing 1-loop Casimir-like contribution (\ref{eq:RenEnergyInt}). 

Finally, the sector \(R>R_0\) is dominated by the classical energy of the twisted kink. As one can see, the 1-loop corrections do 
not change the qualitative behaviour of the classical contribution.

One may worry about the balancing of classical and 1-loop contributions to the energy which results the minimum. 
The energy is formal an expansion in the parameter \(\alpha=\frac{\lambda}{m^2}\)
\begin{equation}\label{eq:formalenexpand}
 E(R)=\frac{1}{\alpha}\epsilon_{-1}(R)+\alpha^0\epsilon_0(R)+\sum_{n=1}\alpha^n\epsilon_n(R).
\end{equation}
The balancing of the terms of order \(\alpha^{-1}\) and \(\alpha^0\) is only valid if the higher loop contributions 
can be neglected. To see more clearly that this is indeed the case for small enough radius \(R\) we expand the field \(\phi\) and rescale
the coordinates \(x^{\mu}\) by
\begin{equation}\label{eq:rescale}
 \phi(x)=\frac{1}{\sqrt{\alpha}}\bar\phi_0+\tilde\phi,\qquad x^{\mu}=R\bar x^{\mu}.
\end{equation}
The fluctuation action is then be written as (using \(\bar\phi_0=0\) for \(R<R_0\))
\begin{equation}
 \tilde S=\int_0^1\mathrm{d}^2\bar x\left[\frac{1}{2}\bar\partial_{\mu}\tilde\phi\bar\partial^{\mu}\tilde\phi-
 +\frac{1}{2}\bar m^2\tilde\phi^2-\frac{1}{4}\bar\lambda\tilde\phi^4\right]
\end{equation}
with the dimensionless parameters
\begin{equation}
 \bar\lambda=R^2\lambda,\qquad \bar m=mR,\qquad \alpha=\frac{\lambda}{m^2}=\frac{\bar\lambda}{\bar m^2}
\end{equation}
in contrast to the case of \(\phi^4\) on the infinite line, where only one dimensionless parameter \(\alpha\) exists.
Now one can see that making the radius of the compactified dimension small (\(R\to 0\)) by holding \(\lambda\) and \(m\) fixed is equivalent to decreasing
the effective dimensionless coupling constant \(\bar\lambda\) for the theory on the unit circle. In the limit \(R\to 0\) one ends up with
a free Gaussian action and the terms of \(\mathcal{O}(\alpha)\) can be neglected. So for some small \(\tilde R<R_0\) the only relevant terms 
are the classical and 1-loop contribution. Since \(R_{min} m\approx\sqrt{\alpha}\) (see (\ref{eq:minradius})), by appropriate choose of \(m\) one
can achieve \(R_{min}<\tilde R\) and the minimum is valid. 

Of course a quantitative estimate of the higher loop effects lies beyond the scope of this paper. Nevertheless a quantitative comparison of semiclassical and 
exact results is in principle possible in the case of the sine-Gordon soliton on \(S^1\) which is an integrable model \cite{Fev2}. The corresponding fluctuation 
equation is the much simpler \(n=1\) Lam\'e equation \cite{Mus3}. This will be discussed elsewhere \cite{Paw2}.

When \(\lambda/m^2>0.88\) the energy of the twisted kink \(E(k)\) becomes negative for certain values of \(k\) (or \(R\)).
Fortunately this happens outside of the region of validity \(\lambda/m^2\ll 1\). Although negative energies are no problem as long as one 
talks about Casimir energies, one runs into troubles if one wants to interpret \(E(k)\) as the mass of the twisted kink. It is known from 
the sine-Gordon model that for values of the coupling constant where the mass of the quantum soliton is formally negative the theory has 
no stable ground state \cite{Cole}. 

\section{Conclusion}
We have constructed an integral representation of the 1-loop energy contribution of the twisted kink of \(\phi^4\)-theory in semiclassical quantization 
appropriate for numerical evaluations. We used special finite-gap properties of the fluctuation equation, which is the \(n=2\) Lam\'e equation, to obtain
an analytic expression for the spectral discriminant \(\Delta(\lambda)\) and the related quasi-momentum \(p(\lambda)\).
Although the Lam\'e equation is a classic subject in mathematical physics, an explicit expression for the quasi-momentum 
in the case \(n=2\) as function of the eigenvalue parameter \(\lambda\), was still missing \cite{Bra2}. 

Our renormalized expressions of the 1-loop energy refine some previously obtained results in \cite{Bra1,Bra2} and completes the discussion
in \cite{Muss} for twisted \(\phi^4\)-theory in the sector \(R>R_0\). We have shown that the large mass renormalization condition cannot
be applied for \(R<R_0\) because the analytic continuations by binomial expansion and integral representation are only valid for \(mR<\pi\).

Therefore we have fixed the energy renormalization in this sector after renormalization for \(R>R_0\) by the condition that the renormalized energies
have to be continuous at \(R=R_0\). In the limit \(R\to\infty\) we have obtained the well known 1-loop energy of the standard \(\phi^4\)-kink.
A further observation we made is, that for \(R<R_0\) a dynamical preferred radius \(R_{min}\) exist. The existence of a minimum in the energy
is a result of the interplay between the classical and 1-loop energy contributions.

From the viewpoint of identifying a mechanism for stabilizing extra dimensions one has to include gravity in our considerations and examine whether minima
in the energy density appear \cite{Gree}. Also one can impose on (\ref{eq:GenSolution}) other boundary conditions as (anti-)periodic, e.g. orbifold 
compactification \(S^1/Z_2\) and analyse the 1-loop effects to the corresponding orbifold kinks \cite{Grza, Cho, Toh1, Toh2}. The fluctuation equation 
remains the same, only the allowed eigenvalues will change by the different boundary conditions.

In order to find the quasi momentum and the spectral discriminant for \(n>2\)  Lam\'e equations, one has to solve a set of \(n\) Bethe ansatz equations, 
which become increasingly complicated for larger \(n\) \cite{Whit}. At least the first derivatives of the quasi momentum up to \(n=5\) 
are known explicitly \cite{Gros}. These Lam\'e equations can be understood as fluctuation equations of kink solutions of some quantum field theory on \(S^1\) 
where the interaction \(V(\phi)\) is only known implicitly. In the limit \(R\to\infty\) the \(n>2\) Lam\'e equations will then lead to the reflectionsless 
potentials considered in \cite{Bor3, Boya}.

\appendix
\section{Analytical continuation by Binomial expansion}
Consider the spectral zeta function (\ref{eq:Regzeta})
\begin{equation}\label{eq:ApZeta}
 \zeta_D(s)=2\mu^{1+2s}\sum_{n=0}^{\infty}\left[\left(\frac{(2n+1)\pi}{R}\right)^2-m^2\right]^{-s}.
\end{equation}
For \(\frac{mR}{\pi}<1\) we can expand (\ref{eq:ApZeta}) in binomials:
\begin{equation}
 \fl \zeta_D(s)=2\mu\left(\frac{R\mu}{\pi}\right)^{2s}\sum_{n=0}^{\infty}\sum_{k=0}^{\infty}\frac{\Gamma(1-s)}{k!\Gamma(1-s-k)}(-1)^k
 \left(\frac{Rm}{\pi}\right)^{2k}(2n+1)^{-2s-2k}.
\end{equation}
In the region of absolute convergence, we can interchange the order of summation:
\begin{equation}
 \zeta_D(s)=2\mu\left(\frac{R\mu}{2\pi}\right)^{2s}\sum_{k=0}^{\infty}\frac{\Gamma(s+k)}{k!\Gamma(s)}\left(\frac{mR}{2\pi}\right)^{2k}
 (2^{2s+2k}-1)\zeta(2s+2k),
\end{equation}
where we have used the following identities:
\begin{equation}
 \frac{\Gamma(1-s)}{\Gamma(1-s-k)}=(-1)^k\frac{\Gamma(s+k)}{\Gamma(s)}
\end{equation}
and
\begin{equation}
 \sum_{n=0}^{\infty}(2n+1)^{-2s-2k}=(1-2^{-2s-2k})\zeta(2s+2k)
\end{equation}
with the Riemann zeta function \(\zeta(s)\).

\section{Spectral discriminant in the limit \(k\to 0\)}
From (\ref{eq:Translos}) one sees, that \(\alpha_1\to i\mathbf{K}'\) for \(k\to 0\). We use therefore the following identities for
Jacobi zeta function with imaginary argument:
\begin{eqnarray}\label{eq:Imaginaer}
 Z(iu,k)&=&i\left[\frac{\mathrm{sn}(u,k')\mathrm{dn}(u,k')}{\mathrm{cn}(u,k')}-Z(u,k')-\frac{\pi u}{2\mathbf{KK'}}\right]=\nonumber\\
 &=& i\left[-i\frac{\mathrm{sn}(iu,k)\mathrm{dn}(iu,k)}{\mathrm{cn}(iu,k)}-Z(u,k')-\frac{\pi u}{2\mathbf{KK'}}\right].
\end{eqnarray}
From (\ref{eq:Translos}) one gets also
\begin{eqnarray}
 \mathrm{cn}^2\alpha_1&=&\frac{2k^2-4+\lambda}{6k^2}-\frac{1}{2k^2}\sqrt{g_2(k)-\frac{1}{3}(\lambda-2(1+k^2))^2},\\
 \mathrm{dn}^2\alpha_1&=&\frac{2-4k^2+\lambda}{6}-\frac{1}{2}\sqrt{g_2(k)-\frac{1}{3}(\lambda-2(1+k^2))^2}.
\end{eqnarray}
So (\ref{eq:Imaginaer}) can be written as
\begin{eqnarray}
  Z(iu,k)=i\left[-i\sqrt{\frac{\frac{4(1+k^2)-\lambda}{6k^2}+\frac{1}{2k^2}\sqrt{g_2(k)-\frac{1}{3}(\lambda-2(1+k^2))^2}}
 {\frac{2k^2-4+\lambda}{6k^2}-\frac{1}{2k^2}\sqrt{g_2(k)-\frac{1}{3}(\lambda-2(1+k^2))^2}}}\times\right.\nonumber\\ \nonumber\\
 \fl \times\left.\sqrt{\frac{2-4k^2+\lambda}{6}-\frac{1}{2}\sqrt{g_2(k)-\frac{1}{3}(\lambda-2(1+k^2))^2}}-Z(u,k')-\frac{\pi u}{2\mathbf{KK}'}\right].
\end{eqnarray}
For \(k\to 0\) and \(u\to\mathbf{K}'\) this reduces to
\begin{equation}
 Z(iu,k)\to i\left[\sqrt{\frac{2+\lambda}{6}-\frac{1}{2}\sqrt{\frac{\lambda(4-\lambda)}{3}}}-1\right].
\end{equation}
The quasi-momentum becomes
\begin{eqnarray}
\fl p(\lambda)\stackrel{k\to 0}{\longrightarrow} -\sqrt{\frac{2+\lambda}{6}-\frac{1}{2}\sqrt{\frac{\lambda(4-\lambda)}{3}}}-
 \sqrt{\frac{2+\lambda}{6}+\frac{1}{2}\sqrt{\frac{\lambda(4-\lambda)}{3}}}+2 =\nonumber\\
\fl =-\sqrt{\frac{2+\lambda}{3}+2\sqrt{\left(\frac{2+\lambda}{6}\right)^2-\frac{1}{4}
 \frac{\lambda(4-\lambda)}{3}}}+2=-\sqrt{\lambda}+2.
\end{eqnarray}
Finally the spectral discriminant becomes
\begin{equation}
 \Delta(\lambda)\to \cos^2\left(\frac{\pi}{2}\sqrt{\lambda}\right).
\end{equation}

\end{document}